\documentclass[aps,pre,superscriptaddress,twocolumn,longbibliography]{revtex4-2}

\usepackage{epsfig}
\usepackage{amsmath,amssymb,bm}
\usepackage{graphicx}
\usepackage{verbatim}
\usepackage{enumerate}

\def\ER{Erd\H{o}s-R\'enyi }

\begin{document}

\title{Exotic Critical Behavior of Weak Multiplex Percolation}

\author{G.~J.~Baxter}
\affiliation{Department of Physics, University of Aveiro $\&$ I3N, Campus Universit\'ario de Santiago, 3810-193 Aveiro, Portugal}

\author{R.~A.~da~Costa}
\affiliation{Department of Physics, University of Aveiro $\&$ I3N, Campus Universit\'ario de Santiago, 3810-193 Aveiro, Portugal}

\author{S.~N. Dorogovtsev}
\affiliation{Department of Physics, University of Aveiro $\&$ I3N, Campus Universit\'ario de Santiago, 3810-193 Aveiro, Portugal}
%\affiliation{A.F. Ioffe Physico-Technical Institute, 194021 St. Petersburg, Russia}

\author{J.~F.~F. Mendes}
\affiliation{Department of Physics, University of Aveiro $\&$ I3N, Campus Universit\'ario de Santiago, 3810-193 Aveiro, Portugal}

\date{\today}

\begin{abstract}
We describe the critical behavior of weak multiplex percolation, a generalization of percolation to multiplex or interdependent networks. A node can determine its active or inactive status simply by referencing neighboring nodes.
This is not the case for the more commonly studied generalization of percolation to multiplex networks, the mutually connected clusters, which requires an interconnecting path within each layer between any two vertices in the giant mutually connected component. 
We study the emergence of a giant connected component of active nodes under the weak percolation rule, finding several non-typical phenomena. In two layers, the giant component emerges with a continuos phase transition, but with quadratic growth above the critical threshold. In three or more layers, a discontinuous hybrid transition occurs, similar to that found in the giant mutually connected component. In networks with asymptotically powerlaw degree distributions, defined by the decay exponent $\gamma$, the discontinuity vanishes  but at $\gamma=1.5$ in three layers, more generally at $\gamma = 1+ 1/(M-1)$ in $M$ layers.
\end{abstract}

%\keywords{}

\maketitle

\section{Introduction \label{intro}}

Complex systems with interdependent sub-systems may be modeled as a multiplex (or colored) network, where links of different types (colors) represent connections within different sub-systems, while nodes having more than one type of connection encompass interdependencies between subsystems \cite{son2012percolation}. Equivalently, one may use a multiplex network, with a layer for each sub-system and links between nodes in different layers representing interdependencies \cite{buldyrev2010catastrophic,son2012percolation}.
To study the resilience of such systems, one typically considers a generalization of percolation. The concept of connected cluster in a single-layer network generalizes to mutually connected clusters, defined as a set of nodes each pair of which is connected by at least one path in all of the layers in which they participate \cite{buldyrev2010catastrophic, baxter2012avalanche, baxter2016unified}. The interdependency between layers (colors) leads to increased fragility of the system, and under random damage, the giant mutually connected component collapses discontinuously \cite{buldyrev2010catastrophic} in a hybrid phase transition of the $k$-core type \cite{baxter2012avalanche,dgm2006}. The collapse occurs due to long range cascading failures in the system \cite{baxter2012avalanche}.

This percolation process applies a global condition to identify surviving nodes: a path of every color must exist between every pair of nodes in a mutually connected cluster in order for the members of the cluster to survive. One may identify the mutually connected clusters by a global pruning process, iteratively removing clusters in each layer that do not have a counterpart cluster in each other layer, until a stable equilibrium is reached. Alternatively, one may identify the connected clusters in each layer, and remove any non-overlapping parts, then repeating the process until no more nodes are pruned.
This percolation process has been extensively studied, with works considering effects of
partial interdependence \cite{dong2012percolation}
overlapping edges \cite{min2015link,baxter2016correlated,cellai2016message}, multiple dependencies \cite{shao2011cascade} correlations \cite{hu2013percolation} among many others \cite{bianconi2018multilayer,kivela2014multilayer,boccaletti2014structure,cozzo2018multiplex}.
As in many network processes, heavy-tailed degree distributions have a strong effect on the phase transition, with the point at which the giant mutually connected cluster emerges $p_c \to 0$ (where $1-p$ is the fraction of nodes initially damaged) as the powerlaw exponent $\gamma \to 2$. Approaching the same point, the size of the discontinuity decays rapidly as $S_c 
\sim 4^{-1/(\gamma-2)}/(\gamma-2)$ \cite{baxter2012avalanche}.

An alternative definition for percolation on multiplex networks was proposed in Ref. \cite{baxter2014weak}, in which the survival of a node is established by a strictly local rule: if it has at least one connection to another surviving node in every layer in which it participates.  In this rule the survival of an agent depends only on its immediate neighborhood.

%%%%%%%%%%%%%%%%%%%%%%%%%%%%%%%%%%%%%%%%%%%%
%%%%%%%%%%%%%%%%%%%%%%%%%%%%%%%%%%%%%%%%%%%%
%%
\begin{figure}[t]
\begin{center}
\includegraphics[width=0.7\columnwidth]{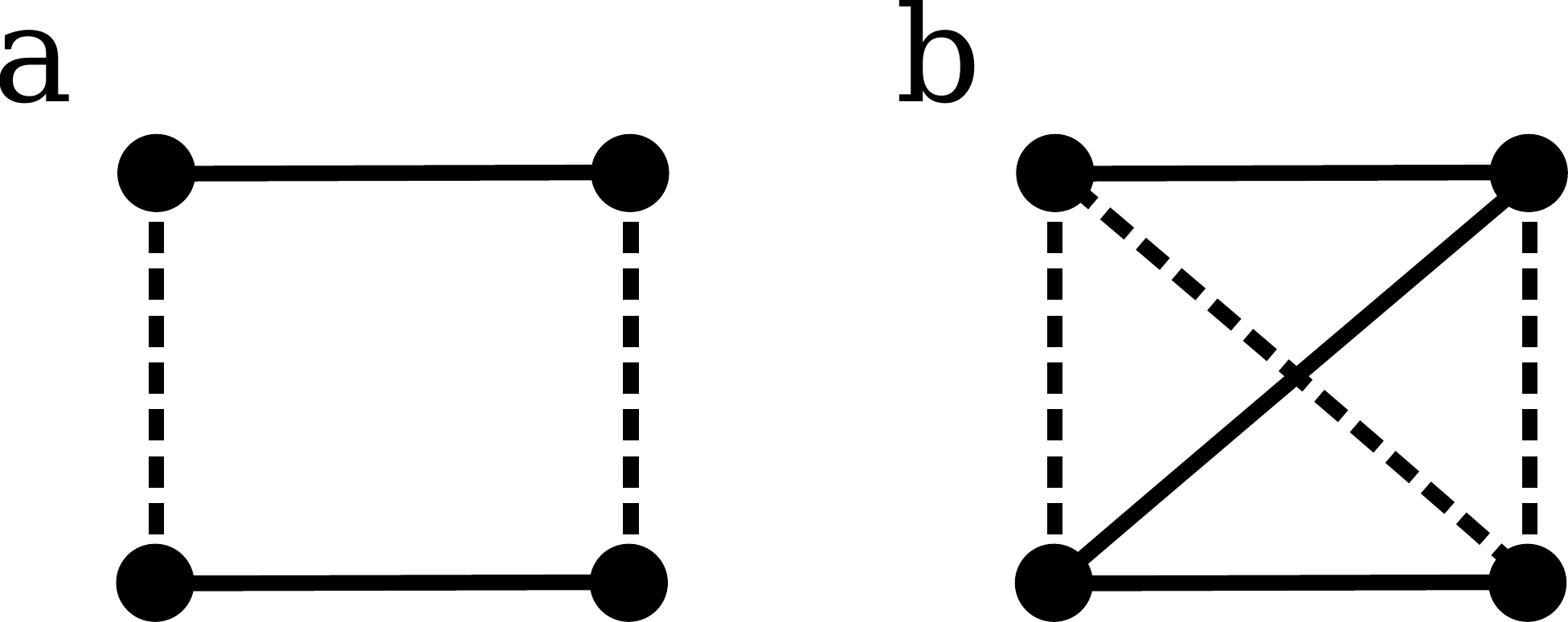}
\end{center}
\caption{Illustration of the difference between weak percolation and mutually connected clusters. In (a) all nodes have connections of both solid and dashed types, and belong to the same weak percolating cluster. This graph contains no mutually connected clusters. (b) In order for a set of nodes to form a mutually connected cluster, there must be a path of both kinds between every pair of vertices. The four vertices shows form a mutually connected cluster and a weak percolating cluster.}
\label{f1}       
\end{figure}
%%
%%%%%%%%%%%%%%%%%%%%%%%%%%%%%%%%%%%%%%%%%%%%
%%%%%%%%%%%%%%%

Due to the less restrictive definition, we refer to this percolation process as weak multiplex percolation. Despite its purely local (and hence more realistic) process, it nevertheless may undergo the same discontinuous hybrid transition as found in the `strong' rule described above.
The phase diagram in networks with rapidly decaying degree distributions was delineated in Ref. \cite{baxter2014weak}.
This process was further explored in Ref. \cite{min2014multiple}, and the relationship with the stronger rule elaborated in Ref.\cite{baxter2016unified}. In $(M\geq2)$-layer networks the problem is equivalent to $(1{-}1{-}...{-}1)$-core percolation, as proposed in Ref. \cite{azimi2014k}.

Here we give the complete critical behavior of this multiplex percolation process in detail. It produces either a continuous second-order transition, with unusual beta-exponent (giving the growth rate of the order parameter above the critical point) of two, or a discontinuous hybrid phase transition, with square-root scaling above the critical point.
We show that heavy-tailed degree distributions, as one might expect, have a strong effect, but in an unusual way. In powerlaw degree distributed networks, with powerlaw exponent $\gamma$, the discontinuous transition disappears at $\gamma = 1.5$ in three layer networks (in general, at $1 + 1/(M-1)$ in $M$ layres) in contrast to the limit $\gamma=2$ found for example in ordinary percolation \cite{cohen2002percolation,dorogovtsev2008critical} and the mutually connected cluster \cite{baxter2012avalanche}.

The remainder of this paper is organized in the following way. In Section \ref{analysis} we define the problem and give the general self-consistency equations which allow for a complete solution. We consider the continuous transition which occurs in two layer networks in Section \ref{s4}. In Section \ref{s5} we consider the discontinuous hybrid transition which occurs in three or more layers. Finally discussion and conclusions are given in Section \ref{conclusions}.

%===========================================================================
%===========================================================================
%===========================================================================
%===========================================================================

\section{Problem and General Analysis\label{analysis}}

%%%%%%%%%%%%%%%%%%%%%%%%%%%%%%%%%%%%%%%%%%%%
%%%%%%%%%%%%%%%%%%%%%%%%%%%%%%%%%%%%%%%%%%%%
%%
\begin{figure}[h]
\begin{center}
\includegraphics[width=\columnwidth]{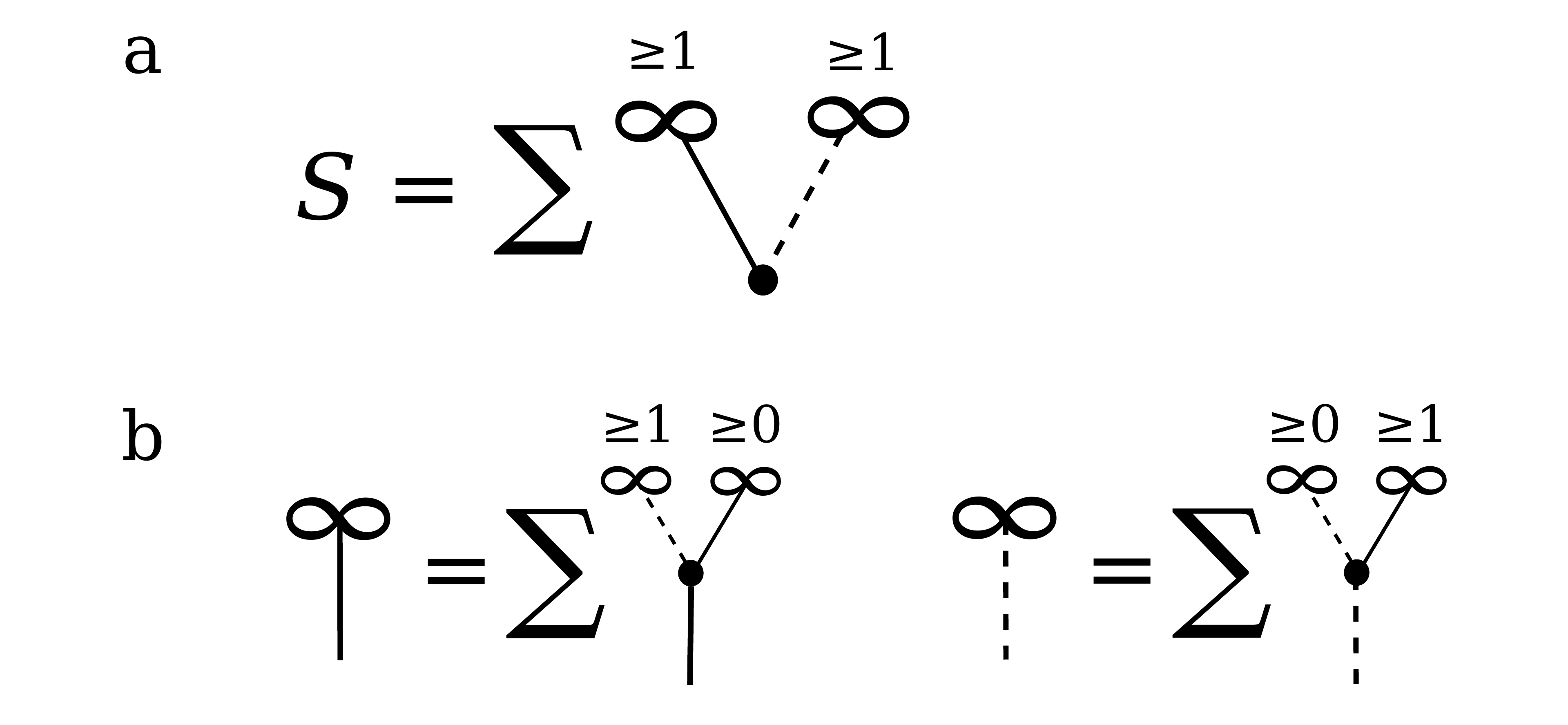}
\end{center}
\caption{
Diagrammatic representation of self-consistency equations for $M=2$ layers, $a$ and $b$.  
(a) In a tree-like network, a node belongs to the giant weak-percolation cluster (giant component) if it has at least one connection via an edge of type $a$ to an infinite subtree satisfying the property $Z_a$ (represented by a solid edge leading to an infinity symbol), and one of type $b$ leading to a satisfying $Z_b$ (dashed edge leading to an infinity symbol), see Eq. (\ref{270}).
 (b) The recursive relations obeyed by the probabilities $Z_a$ (left) and $Z_b$ (right), see Eq. (\ref{260}). For $Z_a$ an edge in layer $a$ leads to a node with at least one edge of type $b$ satisfying $Z_b$, and may also have connections satisfying $Z_a$, and similarly for $Z_b$ , exchanging the labels.}
\label{f4}       
\end{figure}
%%
%%%%%%%%%%%%%%%%%%%%%%%%%%%%%%%%%%%%%%%%%%%%
%%%%%%%%%%%%%%%%%%%%%%%%%%%%%%%%%%%%%%%%%%%%

Let us consider a large sparse random multiplex network, consisting of $N$ nodes connected in one or more of $M$ layers (each having its own unique type of edge). Note that a node does not necessarily participate in all layers. The analysis which follows is not impeded by the presence of degree correlations between layers, so we consider a generalised configuration model network defined by its joint degree distribution $P(q_1,q_2,...,q_M)$. 
A node may be considered active if it retains at least one connection to other active nodes in each of the layers in which it participates. A weak percolating cluster is then a set of such active nodes which are connected to each other (each member is connected to at least one other member in at least one layer).

In the large size limit $N\to\infty$ we can use the locally tree-like property of the network to write self-consistency equations for the probability that a randomly selected node is active.  This is equal to the the relative size of the giant weak-percolation cluster (we will from now on use ``giant component" as a shorthand for this cluster) in such networks. 
The size of giant component in an infinite sparse network is given by
\begin{align}\label{S_general}
S = 
\sum_{q_1,q_2,...,q_M}\!\!\!\!\!\!
P(q_1,q_2,...,q_M)
\prod_{\alpha = 1}^M
[1-(1-Z_{\alpha})^{q_{\alpha}}]
\end{align}
where the probabilities $Z_1, Z_2, ..., Z_M$ are given by 
\begin{align}\label{Z_general}
Z_{\alpha} = 
\sum_{q_1,q_2,...,q_M}\!\!\!\!\!\!
\frac{q_{\alpha} P(q_1,q_2,...,q_M)}{\langle q_{\alpha}\rangle}
\prod_{\beta \neq {\alpha}}
[1-(1-Z_{\beta})^{q_{\beta}}]
\end{align}
for ${\alpha} = 1,2,...,M$. They represent the probability that, upon following an edge of type ${\alpha}$,
we encounter a vertex with at least one edge of type $\beta$ satisfying $Z_\beta$ for all layers $\beta \neq \alpha$. These equations are illustrated diagrammatically for two layers in Fig. \ref{f4}.

One may then obtain the size of the giant component by solving, 
Eqs. (\ref{Z_general}) and substituting the solution into Eq. (\ref{S_general}).
In a two layer network with rapidly decaying degree distribution (such as an \ER network), as connectivity increases, the giant component appears continuously with a second-order phase transition, which differs from the standard percolation transition as the giant component grows quadratically above the critical threshold.
For three or more layers, on the other hand, one finds that the giant component appears with a discontinuous hybrid transition, similar to that seen in $k$-core percolation. The size of the giant component $S$ jumps from zero to a finite value at a critical threshold, and then grows as the square root of the distance above the threshold.

%===========================================================================
%===========================================================================
%===========================================================================
%===========================================================================

\section{Two layers - continuous transition}
\label{s4} 

Let us first consider the case of two layers. In this case the giant percolating cluster emerges with a continuous transition, but with different characteristics than the ordinary percolation transition.

Equation (\ref{Z_general}) becomes
\begin{eqnarray}
Z_a &=& \sum_{q_a,q_b} \frac{q_a P(q_a,q_b)}{\langle q_a \rangle} 
\,1\,[1-(1-Z_b)^{q_b}]
,  
\nonumber
\\[3pt]
Z_b &=& \sum_{q_a,q_b} \frac{q_b P(q_a,q_b)}{\langle q_b \rangle} 
[1-(1-Z_a)^{q_a}]\,
\label{260}
\end{eqnarray}
and the expression for the size of the giant weak percolation cluster (which corresponds to the $(1{-}1)$-core multiplex $k$-core) is 
\begin{equation}
S
= \sum_{q_a,q_b} P(q_a,q_b) 
[1-(1-Z_a)^{q_a}][1-(1-Z_b)^{q_b}]
.  
\label{270}
\end{equation}

\subsection{Rapidly decaying degree distributions}

If the moments $\langle q_a \rangle$, $\langle q_b \rangle$, $\langle q_a q_b\rangle$, $\langle q_a^2 q_b \rangle$, and $\langle q_a q_b^2 \rangle$ are finite, which is the case, for example, for \ER network layers,
we may expand Eqs. (\ref{260}) for small $Z_a$ and $Z_b$, finding
\begin{eqnarray}
Z_a &\cong&  \frac{1}{\langle q_a \rangle} 
\bigl[
\langle q_a q_b \rangle Z_b - \frac12 \langle q_a q_b (q_b -1 ) \rangle Z_b^2
\bigr]
,  
\nonumber
\\[3pt]
Z_b &\cong&  \frac{1}{\langle q_b \rangle} 
\bigl[
\langle q_a q_b \rangle Z_a - \frac12 \langle q_a q_b (q_a -1 ) \rangle Z_a^2
\bigr]
\label{272}
\end{eqnarray}
which indicates that the continuous transition occurs when
\begin{equation}
\langle q_a \rangle \langle q_b \rangle =  \langle q_a q_b \rangle^2\,.  
\label{276}
\end{equation}

Near the transition point, we can write
\begin{equation}
S
\cong \langle q_a q_b \rangle Z_a Z_b  
,  
\label{274}
\end{equation}
and using Eq.~(\ref{272}) we find the size of the giant cluster near the critical point: 
\begin{eqnarray}
S &\cong& 
\frac{4\,{\langle q_a \rangle \langle q_b \rangle}\bigl( \langle q_a q_b \rangle^2 - \langle q_a \rangle \langle q_b \rangle \bigr)^2}
{
{\langle q_a q_b \rangle}
Q_a
Q_b
}
.
\label{278}
\end{eqnarray}
where for compactness we have defined
\begin{align}
Q_a \equiv \bigl[  \langle q_a q_b(q_a {-} 1) \rangle \langle q_b \rangle + \langle q_a q_b(q_b {-} 1) \rangle \langle q_ a q_b \rangle \bigr],\\
Q_b \equiv \bigl[  \langle q_a q_b(q_b {-} 1) \rangle \langle q_a \rangle + \langle q_a q_b(q_a {-} 1) \rangle \langle q_ a q_b \rangle \bigr].
\end{align}

The giant weakly percolating component grows as the square of the distance from the critical point, i.e. $\beta = 2$, as opposed to the usual percolation transition which has $\beta = 1$. 
To illustrate this, let us consider the simplified case in which a node's degrees in each layer are independent, $P(q_a,q_b)=P_a(q_a)P_b(q_b)$.  
Then $\langle q_aq_b\rangle = \langle q_a\rangle \langle q_b\rangle$, and the condition of Eq. (\ref{276}) may be reduced to
(assuming still that $\langle q_a \rangle, \langle q_b \rangle < \infty$)
\begin{equation}
\langle q_a \rangle \langle q_b \rangle = 1
.  
\label{292}
\end{equation}
Then near the transition, 
\begin{eqnarray}
&& 
\!\!\!\!\!
S = Z_aZ_b \cong 
\frac{4\langle q_a \rangle^3(\langle q_a \rangle\langle q_b \rangle - 1)^2}
{[(\langle q_b^2 \rangle \langle q_a \rangle^2 {-} 1)\langle q_a \rangle + \langle q_a^2 \rangle - \langle q_a \rangle]^2  
} 
,  
\label{296}
\end{eqnarray}
where we have used that $\langle q_b \rangle = 1/\langle q_a \rangle$ at the threshold. 
Further, if the network is symmetric ($\langle q_a\rangle = \langle q_b\rangle \equiv \langle q\rangle$), then the giant component exists for $\langle q \rangle >1$. Assuming, moreover, $\langle q^2 \rangle$ is finite, we arrive at the following relative size of the giant component near the transition
\begin{equation}
S \cong 4\, \frac{(\langle q \rangle-1)^2}{(\langle q^2 \rangle-1)^2} 
.  
\label{298}
\end{equation}
Here $(\langle q \rangle-1)$ plays the role of a control parameter, and we see immediately that the growth is quadratic.
In the symmetric Erd\H{o}s--R\'enyi situation, $S$ coincides with the square of the relative size of the giant connected component in an individual layer. 

%%%%%%%%%%%%%%%%%%%%%%%%%%%%%%%%%%%%%%%%
\subsection{Heavy-tailed degree distributions}

For strongly heterogeneous degree distributions the condition that the leading moments are finite may not be met. 
In this case we may use generating functions to study the asymptotics of the solutions. 
When $P(q_a,q_b) = P_a(q_a)P_b(q_b)$, we may rewrite Eqs.~(\ref{260}) using generating functions (see Appendix \ref{sa2}) as 
\begin{eqnarray}
Z_a &=& 1 - G_b(1-Z_b)
,  
\nonumber
\\[3pt]
Z_b &=& 1 - G_a(1-Z_a) 
.
\label{300}
\end{eqnarray}
while the size of the giant percolating cluster is simply
\begin{equation}
S = Z_aZ_b   
.
\label{301}
\end{equation}

For concreteness, we will consider uncorrelated powerlaw tailed degree distributions of the form 
\begin{align}
P(q) = Aq^{-\gamma}\
\end{align}
for each layer.
Note that, as we will see, the exponent  found above for rapidly decaying degree distributions, $\beta=2$, applies for $\gamma>3$ in contrast  to ordinary percolation (where the limiting exponent $\beta=1$ applies only for $\gamma>4$).

%%%%%%%%%%%%%%%%%%%%%%%%%%%%%%%%%%%%%%%%%%%%%%%%%%%%%%%%%%%%%%%%%%%%%%%%%%%%%%%%%
\begin{figure}
\includegraphics[width=\columnwidth]{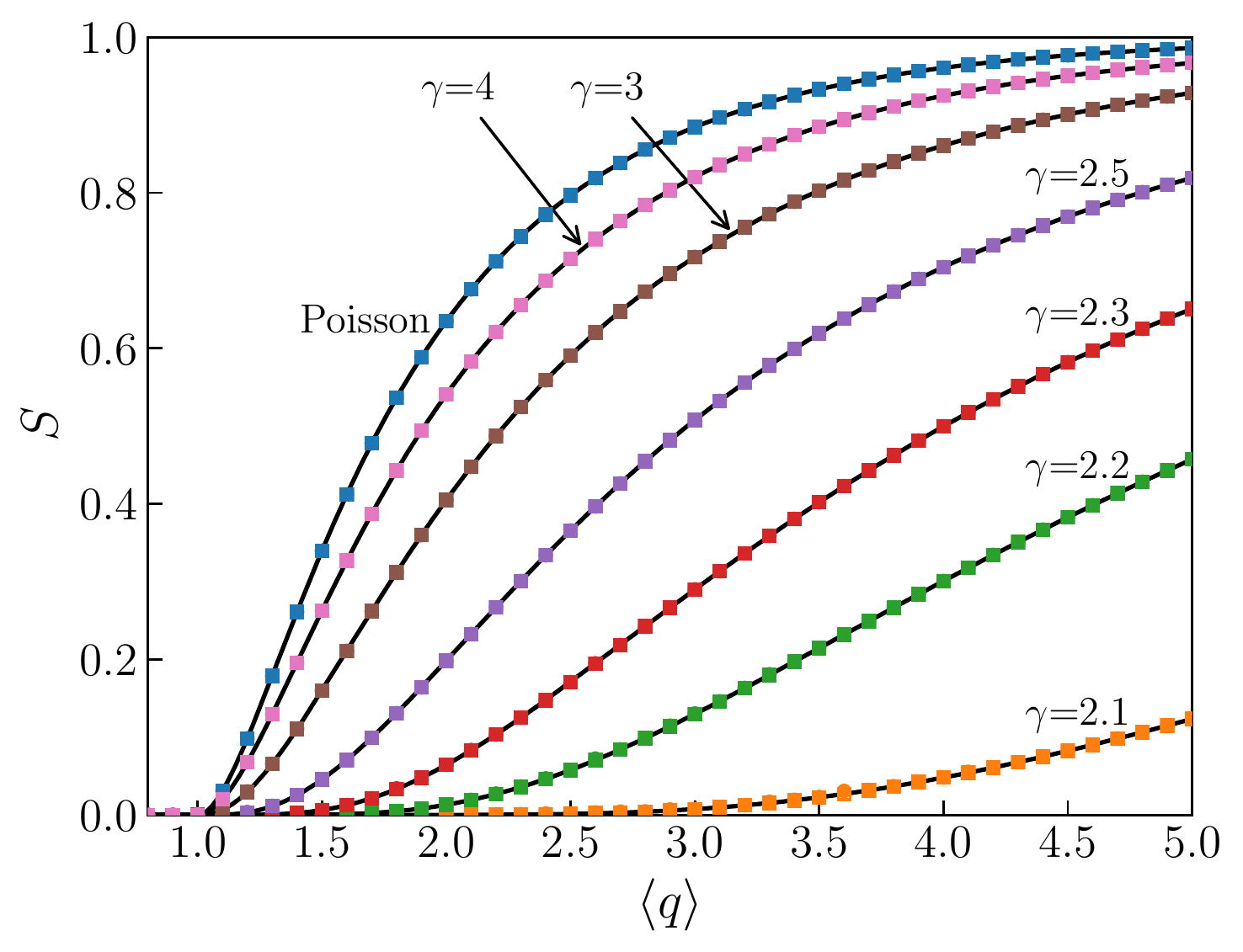}
\caption{Relative size of the largest weak-percolating cluster $S$ as a function of mean degree $\langle q\rangle$ in symmetric two layer multiplex networks, $M = 2$, with each layer having a powerlaw degree distribution $P(q) \sim Aq^{-\gamma}$ generated using the static model, containing  $N=10^6$ and $N=10^7$ nodes (results are virtually identical), see Appendix \ref{simulations}. Results for Poisson degree distributions are shown for comparison. Black solid curves in each case are numerical solutions of Eqs. (\ref{S_general}) and (\ref{Z_general}) .
}\label{fig_M2}
\end{figure}
%%%%%%%%%%%%%%%%%%%%%%%%%%%%%%%%%%%%%%%%%%%%%%%%%%%%%%%%%%%%%%%%%%%%%%%%%%%%%%%%%

Let us first consider the symmetric case, $P(q_a) = P(q_b) \equiv P(q)$.
When $2<\gamma<3$, using the asymptotic behaviour of the generating function, see Appendix \ref{sa2},
\begin{equation}
Z \cong \langle q \rangle Z - A\Gamma(1-\gamma)Z^{\gamma-1}
,  
\label{380}
\end{equation}
so the size of the giant component equals 
\begin{equation}
S = Z^2 \cong \Bigl[\frac{\langle q \rangle-1}{A\Gamma(1-\gamma)}\Bigr]^{2/(\gamma-2)}
.  
\label{400}
\end{equation}
In Fig. \ref{fig_M2} we compare this theoretical calculation of $S$ with simulated networks containing $N=10^6$ and $10^7$ nodes, above the threshold ($c\equiv \langle q\rangle = 1$), showing perfect agreement. 
 Each layer is an independently generated configuration model network, generated according to the static model degree distribution, which is asymptotically powerlaw \cite{catanzaro2005analytic,goh2001universal}. See Appendix \ref{simulations} for more details.
This figure illustrates the very slow growth for values of $\gamma$ close to $2$, and the approach to the quadratic growth at $\gamma\geq 3$. Notice also the lack of significant finite size effects. The results for both network sizes are virtually identical.

When $1<\gamma<2$, the mean degree diverges and one must proceed with caution, as the conditions required for the self consistency equations to be exact may not hold. Nevertheless we find, after extensive comparison against numerical simulations, that our equations give accurate results.
In this region we can no longer use the mean degree as a control parameter. Instead, we may consider applying random damage to the network. 
If edges are retained with probability $p$ and removed with probability $1-p$, the tail of the degree distribution retains the same powerlaw exponent $\gamma$, with a reduced coefficient $A_p = A_1 p^{\gamma-1}$ 
, as we show in Appendix \ref{sa1}.
In this case we find
\begin{equation}
Z \cong  - A_1p^{\gamma-1}\Gamma(1-\gamma)Z^{\gamma-1}
,  
\label{410}
\end{equation}
so, the size of the giant component is 
\begin{equation}
S = Z^2 \cong [-A_1\Gamma(1-\gamma)]^{2/(2-\gamma)} p^{2(\gamma-1)/(2-\gamma)} 
.  
\label{430}
\end{equation}
For site percolation, we have
\begin{equation}
Z \cong  p[- A_1\Gamma(1-\gamma)]Z^{\gamma-1}
,  
\label{410b}
\end{equation}
and so 
\begin{equation}
S = pZ^2 \cong [-A_1\Gamma(1-\gamma)]^{2/(2-\gamma)} p^{(4-\gamma)/(2-\gamma)} 
.  
\label{430b}
\end{equation}
Thus, for both edge and vertex removal, the giant component appears immediately from $p_c = 0$.

We now consider the cases in which the exponents for each layer are different, $P_a(q_a) = Aq_a^{-\gamma_a}$ and $P_b(q_b) = Aq_b^{-\gamma_b}$. In general, the critical behaviour is determined by the smaller of the two degree distribution exponents.
Without loss of generality, let us assume $\gamma_a > \gamma_b$. Results for the opposite case can be obtained by simply exchanging the subscripts $a$ and $b$.
If both exponents are greater than $3$, we have the behavior described in the previous Section.

 We first consider the case  $\gamma_a>3$, $2<\gamma_b<3$. Then $\gamma_a-1 > 2$, so the leading terms in the expansion of $Z_b$ are linear and quadratic. We may neglect the quadratic term, so we have
\begin{eqnarray}
Z_a &\cong& \langle q_b \rangle Z_b - A_b\Gamma(1-\gamma_b)Z_b^{\gamma_b-1}
,  
\nonumber
\\[3pt]
Z_b &\cong& \langle q_a \rangle Z_a  
.
\label{440}
\end{eqnarray}

The solution is 
\begin{equation}
Z_a  \cong \left[\frac{\langle q_a \rangle\langle q_b \rangle - 1}{A_b\Gamma(1-\gamma_b)}\right]^{1/(\gamma_b-2)} \langle q_a \rangle^{-(\gamma_b-1)/(\gamma_b-2)}  
,
\label{450}
\end{equation}
so 
\begin{equation}
S = \left[\frac{\langle q_a \rangle\langle q_b \rangle - 1}{A_b\Gamma(1-\gamma_b)}\right]^{2/(\gamma_b-2)} \langle q_a \rangle^{-\gamma_b/(\gamma_b-2)} 
.  
\label{460}
\end{equation}
Note that, somewhat counterintuitively, the fatter-tailed degree distribution (smaller exponent) determines the behavior, in contrast to what would be the case for more traditional percolation problems. 

If both exponents are less than three, $2<\gamma_a,\gamma_b<3$, we have
\begin{align}
Z_a &\cong \langle q_b \rangle Z_b - A_b\Gamma(1-\gamma_b)Z_b^{\gamma_b-1}
,  
\nonumber
\\
Z_b &\cong \langle q_a \rangle Z_a - A_a\Gamma(1-\gamma_a)Z_a^{\gamma_a-1}  
.
\label{432}
\end{align}
Substituting the second line into the first,
\begin{multline}
(\langle q_a \rangle\langle q_b \rangle - 1)Z_a 
\\
\cong 
\langle q_b \rangle A_a \Gamma(1{-}\gamma_a) Z_a^{\gamma_a-1} 
+
\langle q_a \rangle^{\gamma_b-1} A_b\Gamma(1{-}\gamma_b)Z_a^{\gamma_b-1}.
\label{433}
\end{multline}
If $\gamma_a>\gamma_b$, then the first term on the right-hand side of Eq.~(\ref{433}) should be neglected, and we obtain
\begin{equation}
Z_a \cong \left[\frac{\langle q_a \rangle\langle q_b \rangle - 1}{\langle q_a \rangle^{\gamma_b-1} A_b \Gamma(1{-}\gamma_b)}\right]^{1/(\gamma_b-2)} 
,
\label{436}
\end{equation}
thus
\begin{equation}
S \cong 
\langle q_a \rangle^{-\gamma_b/(\gamma_b-2)} \left[\frac{\langle q_a \rangle\langle q_b \rangle - 1}{ A_b \Gamma(1{-}\gamma_b)}\right]^{2/(\gamma_b-2)}
.  
\label{437}
\end{equation}

Now let us consider $1<\gamma_b<2$, $\gamma_a > 2$. 
We obtain
\begin{align}
Z_a &\cong - A_b\Gamma(1-\gamma_b)Z_b^{\gamma-1}
,  
\nonumber
\\
Z_b &\cong \langle q_a \rangle Z_a  
.
\label{470}
\end{align}
Note that, given the first equation, we can neglect higher order terms in the second equation, so it applies both for $2<\gamma_a<3$ and $\gamma_a>3$, and we treat both these cases together. 

As before, we apply random damage, using the retention probability $p$ as the control parameter.
The degree distribution for layer $b$ is then asymptotically $A_bp^{\gamma_b-1}q^{-\gamma_b}$, where $A_ {b,1}\equiv A_b(p{=}1)$.
The solution to Eqs. (\ref{470}) is 
\begin{align}
\!\!\!\!Z_a  &{\cong} [-A_b\Gamma(1-\gamma_b)]^{1/(\gamma_b-2)} \langle q_a \rangle^{-(\gamma_b-1)/(2-\gamma_b)}  \nonumber\\
& \propto p^{-(\gamma_b-1)/(2-\gamma_b)}
,
\label{480}
\end{align}
so that  
\begin{align}
S &= \langle q_a \rangle Z_a^2  
\nonumber\\
&\cong
[-A_{b,1}\Gamma(1{-}\gamma_b)]^{2/(\gamma_b-2)} \langle q_a \rangle^{-\gamma_b/(2-\gamma_b)} p^{-(\gamma_b-1)/(2-\gamma_b)} 
.  
\label{490}
\end{align}
Note that this expression contains only $\langle q_a \rangle$ of layer $a$. 
Again, the behaviour is determined by the fatter tailed degree distribution.

Finally, when both exponents are small, $1<\gamma_a,\gamma_b <2$, the critical behavior depends on both of them.
Indeed, the equations for $Z_a$ and $Z_b$ have the form: 
\begin{eqnarray}
Z_a &\cong& - A_b\Gamma(1-\gamma_b)Z_b^{\gamma_b-1}
,  
\nonumber\\
Z_b &\cong& - A_a\Gamma(1-\gamma_a)Z_a^{\gamma_a-1}.
\label{492a}
\end{eqnarray}
So 
\begin{align}
S =& Z_aZ_b \nonumber\\
\cong & \Bigl[- A_b\Gamma(1{-}\gamma_b)\Bigr]^{\gamma_a/[1-(\gamma_b-1)(\gamma_a-1)]}\nonumber\\ 
&\times
\Bigl[- A_a\Gamma(1{-}\gamma_a)\Bigr]^{\gamma_b/[1-(\gamma_b-1)(\gamma_a-1)]}
\nonumber\\
\cong &
\Bigl[- A_{a1}\Gamma(1{-}\gamma_a)\Bigr]^{\gamma_b/[1-(\gamma_b-1)(\gamma_a-1)]}\nonumber\\
&\times\Bigl[- A_{b1}\Gamma(1{-}\gamma_b)\Bigr]^{\gamma_a/[1-(\gamma_b-1)(\gamma_a-1)]} \nonumber\\
&\times p_a^{(\gamma_a-1)\gamma_b/[1-(\gamma_b-1)(\gamma_a-1)]} 
p_b^{(\gamma_b-1)\gamma_a/[1-(\gamma_b-1)(\gamma_a-1)]}
.
\label{492aa}
\end{align}
Here we assumed that the fractions of retained edges in layers $a$ and $b$ are $p_a$ and $p_b$ respectively. 

Finally, for completeness, we may make the same calculation for the case of vertex removal. Vertices survive with probability $p$. A factor of $p$ is added to Eq. (\ref{492a}) [compare Eq. (\ref{410b})]. Solving for $Z_a$ and $Z_b$ then substituting into $S = p Z_aZ_b$ we find
\begin{align}
S \cong &\Bigl[- A_{a1}\Gamma(1{-}\gamma_a)\Bigr]^{\gamma_b/[1-(\gamma_b-1)(\gamma_a-1)]}\nonumber\\
&\times\Bigl[- A_{b1}\Gamma(1{-}\gamma_b)\Bigr]^{\gamma_a/[1-(\gamma_b-1)(\gamma_a-1)]} \nonumber\\
&\times p^{(\gamma_a\gamma_b)/[1-(\gamma_b-1)(\gamma_a-1)]}
.
\label{492b}
\end{align}

%===========================================================================
%===========================================================================
%===========================================================================
%===========================================================================

\section{Higher number of layers}
\label{s5}

For more than two layers, the giant weakly percolating component typically appears with a discontinuous hybrid transition \cite{baxter2014weak,baxter2016unified}, of the same type observed in the mutually connected cluster \cite{baxter2012avalanche} and in $k$-core percolation \cite{dgm2006}.

For $M\geq 2$ layers, we have 
\begin{equation}
Z_\alpha = \prod_{\beta\neq\alpha}^M[1 - G_\beta(1-Z_\beta)]
\label{560}
\end{equation}
for $\alpha=1,2,...,M$, and 
\begin{equation}
S = \prod_{\beta=1}^M[1 - G_\beta(1-Z_\beta)] 
= \Bigl(\prod_{\beta=1}^M Z_\beta\Bigr)^{1/(M-1)}
.  
\label{570}
\end{equation}

In the symmetric case, in which every layer is a random network with the same degree distribution, $P(q_a,q_b,q_c, ...)=P(q_a)P(q_b)P(q_c)...$, we have 
\begin{equation}
Z = [1 - G(1-Z)]^{M-1}
\label{580}
\end{equation}
and 
\begin{equation}
S = Z^{M/(M-1)}
.  
\label{590}
\end{equation}

In this situation, for Poisson degree distributions with $c \equiv \langle q \rangle$, Eq.~(\ref{580}) leads to the equation: 
\begin{equation}
Z = [1 - e^{-cZ}]^{M-1}
\label{592}
.
\end{equation}
This equation is practically identical to the one obtained in Ref.~\cite{gao2011robustness} for the relative size $S^\ast$ of the giant mutually connected component in $M$-layer multiplex \ER networks: 
\begin{equation}
S^\ast = [1 - e^{-cS^\ast}]^M
\label{594}
.
\end{equation}
Comparing Eqs.~(\ref{590}) and (\ref{592}) with Eq.~(\ref{594}), we obtain the following relation between these two problems: 
\begin{eqnarray}
S(c,M) &=& S^{\ast\,M/(M-1)}(c,M-1)
,
\nonumber
\\[3pt]
c_\text{c}(M) &=& c^\ast_\text{c}(M-1)
,
\label{596}
\end{eqnarray}
where $M\geq3$, and  $c_c(M)$ and $c_c^\ast(M)$ are the critical value of the average degree for the giant mutually connected component and weak percolation, respectively, for $M$-layer multiplex \ER networks. 
Thus the weak percolation problem on a $M$-layer multiplex \ER network is equivalent to the problem of giant mutually connected component in the corresponding $M-1$-layer multiplex \ER network.

For large $M$, the asymptotics of these quantities are the same in both problems: 
\begin{eqnarray}
c_\text{c} &\cong& \ln M + \ln\ln M + 1 + \frac{\ln\ln M}{\ln M}
,
\nonumber
\\[3pt]
S_\text{c} &\cong& 1 - \frac{1}{\ln M} + \frac{\ln\ln M}{\ln^2 M}
.
\label{598}
\end{eqnarray}
%%

%%%%%%%========================================================================================================================
\subsection{Effect of heavy-tailed degree distributions}

%%%%%%%%%%%%%%%%%%%%%%%%%%%%%%%%%%%%%%%%%%%%
\begin{figure}
\includegraphics[width=\columnwidth]{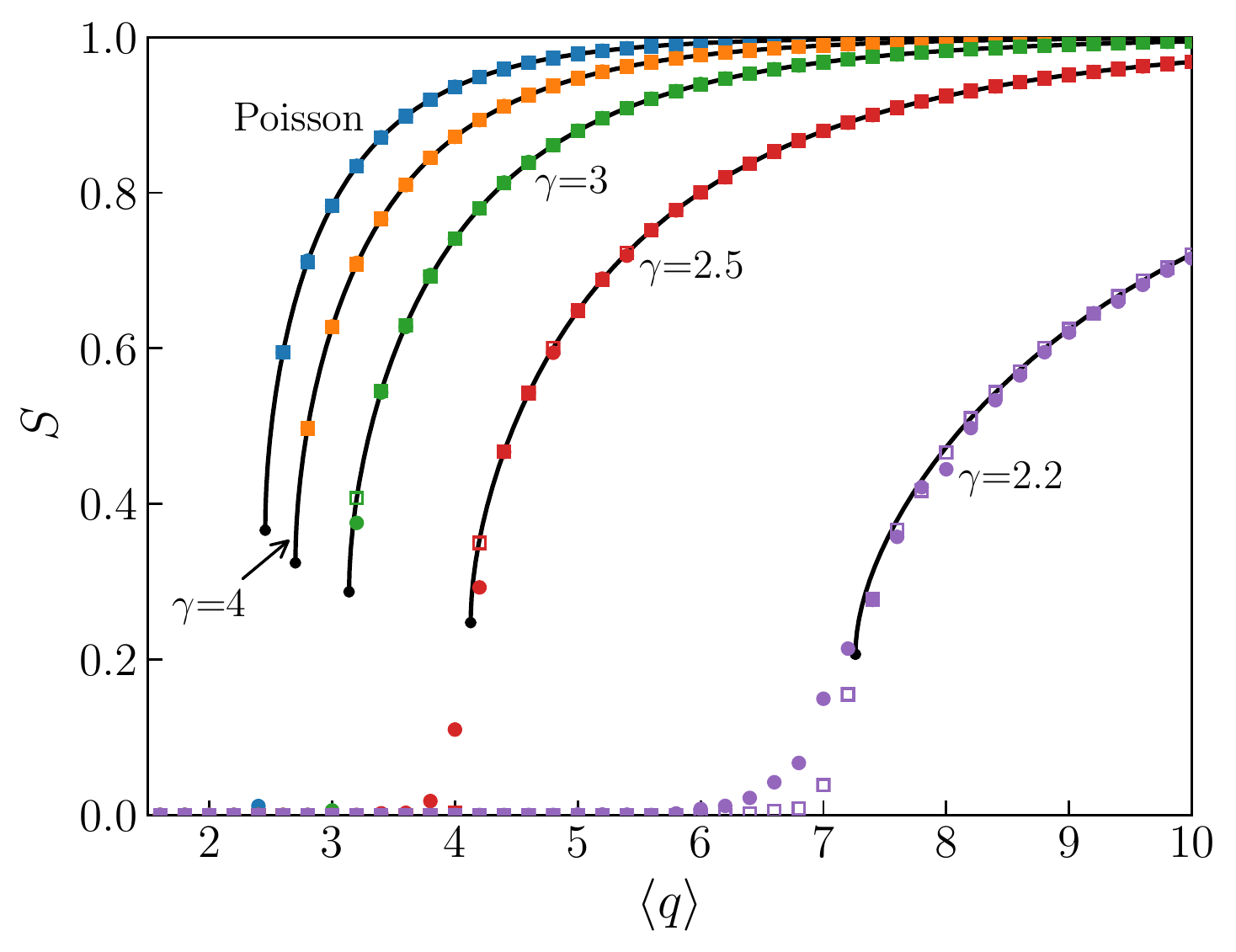}%fig_M3_g_larg_2_v2_nt}%
\caption{Relative size of the giant component $S$ for three identically powerlaw distributed layers ($M=3$) as a function of mean degree $\langle q\rangle$  for various values of $\gamma$  greater than $2$. For details of the simulations see Appendix \ref{simulations}. Black curves show analytic results (from numerical solution of Eqs. (\ref{S_general}) and (\ref{Z_general})), symbols show measurements averaged over 100 synthetic networks of $N=10^4$ nodes (circles) and $N=10^7$ nodes (squares).\label{M3_large}}
\end{figure}
%%%%%%%%%%%%%%%%%%%%%%%%%%%%%%%%%%%%%%%%%%%%

The discontinuous hybrid transition always maintains the same square-root scaling above the transition, however the size of the discontinuity and the location of the critical point for such transitions may be strongly affected by the degree distribution \cite{baxter2012avalanche}. For orientation, we again begin with the symmetric case, $P(q) = Aq^{-\gamma}$. 
As can be seen in Fig. \ref{M3_large}, the location and size of the discontinuity depends strongly on the powerlaw exponent $\gamma$. 
Simulation results were obtained using independently generated configuration model networks following the static model degree distribution for each layer, just as in Fig. \ref{fig_M2}, see Appendix \ref{simulations}.
As $\gamma$ approaches $2$, the mean degree diverges and finite size effects become particularly prominent, as evidenced by the divergence between theoretical and numerical results, and between numerical results for networks of different sizes at $\gamma=2.2$ in the figure.

%%%%%%%%%%%%%%%%%%%%%%%%%%%%%%%%%%%%%%%%%%%%
\begin{figure}
\includegraphics[width=\columnwidth]{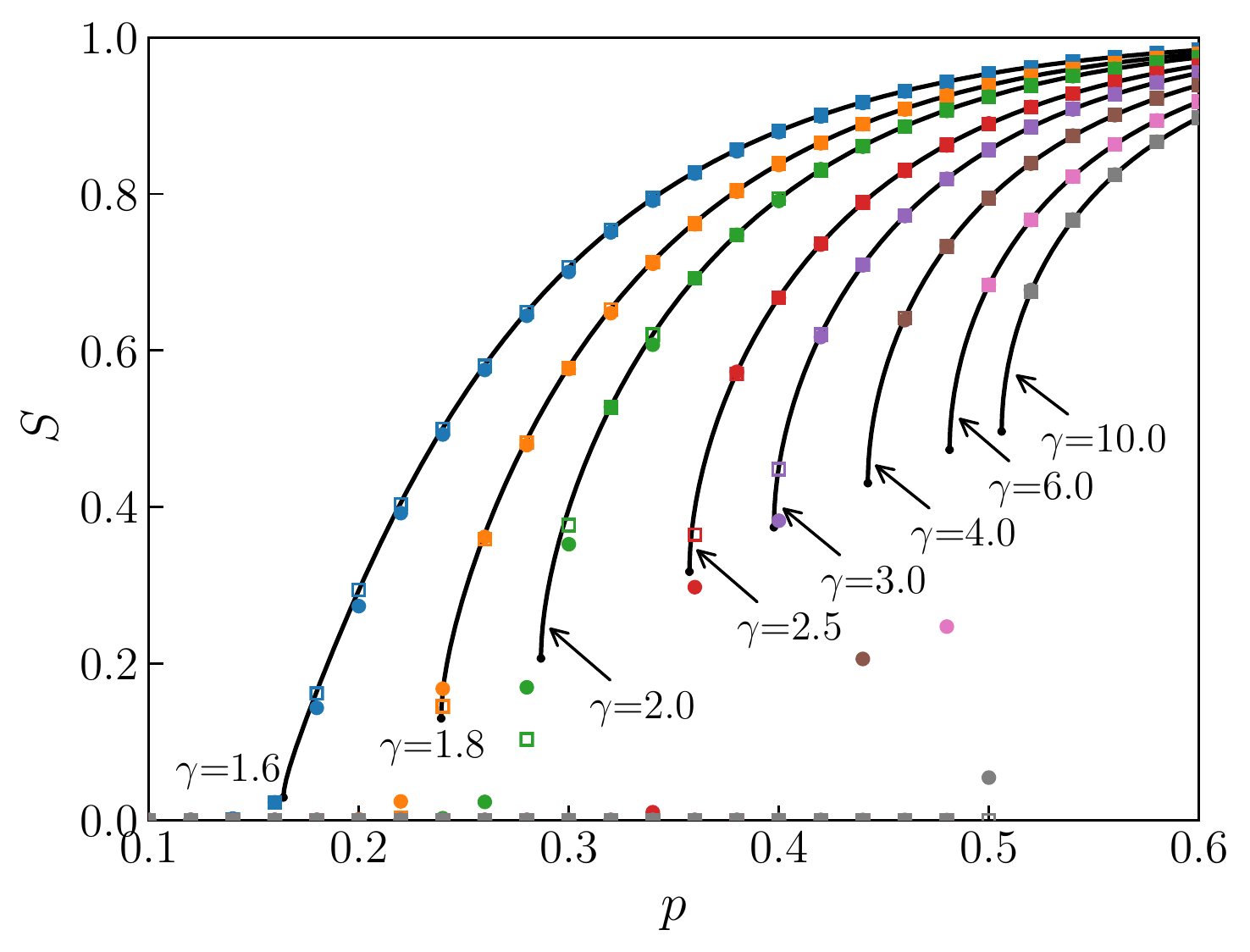}%fig_M3_g_smal_2_larg_15_nt}%
\caption{Relative size of the giant component $S$ for three identically powerlaw distributed layers ($P(q) =Aq^{-\gamma}, q \geq 4$) as a function of undamaged fraction $p$ for various values of $\gamma$  greater than $1.5$. Black curves show analytic results, symbols show measurements for synthetic networks of $N=10^5$ nodes (circles, 100 realisations) and $N=10^7$ nodes (squares, one realisation), see Appendix \ref{simulations}.
\label{M3_small2}}
\end{figure}
%%%%%%%%%%%%%%%%%%%%%%%%%%%%%%%%%%%%%%%%%%%%

%%%%%%%%%%%%%%%%%%%%%%%%%%%%%%%%%%%%%%%%%%%%
\begin{figure}
\includegraphics[width=\columnwidth]{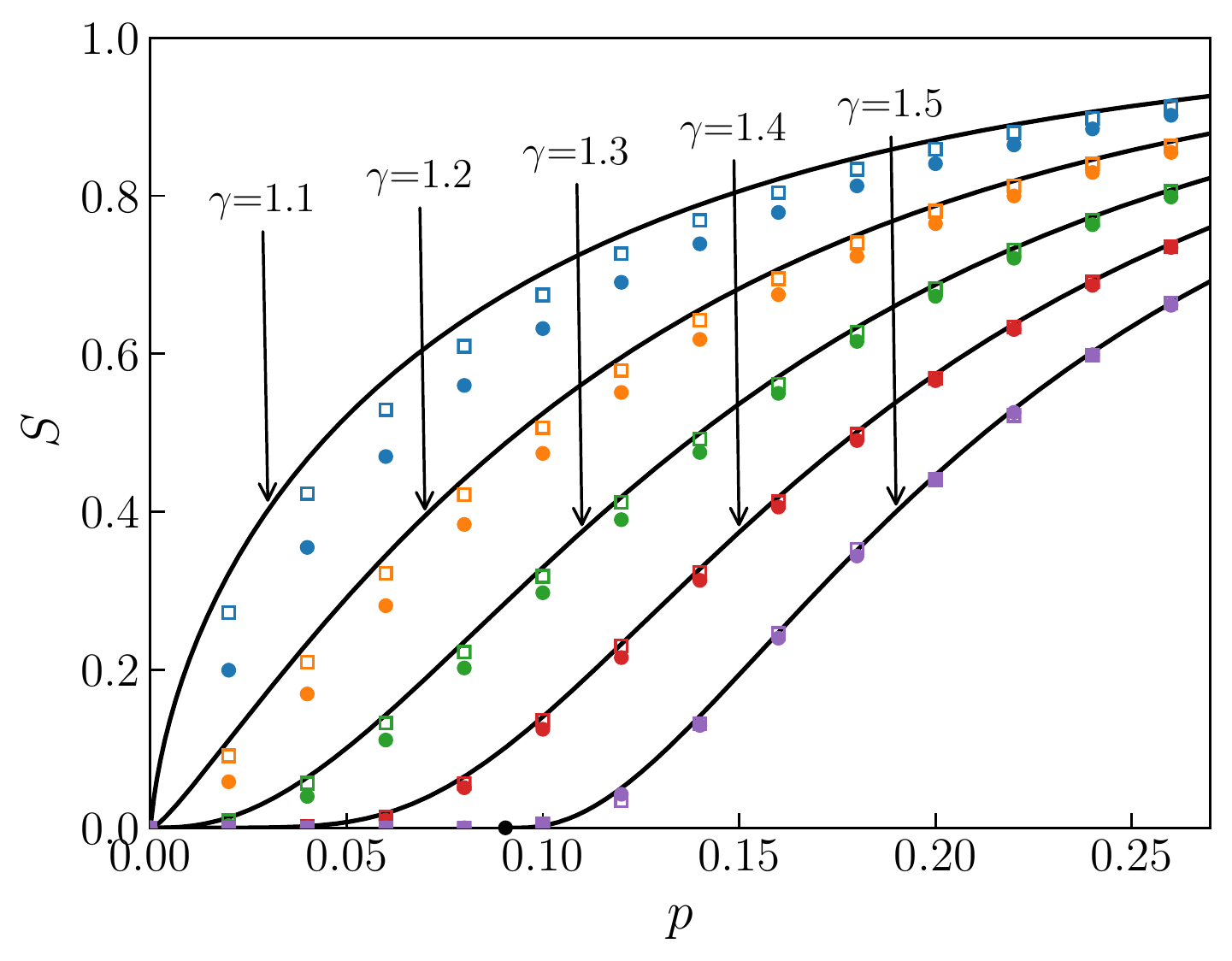}%fig_M3_g_smal_15_nt}%
\caption{Relative size of the giant component $S$ for three identically powerlaw distributed layers ($P(q) =Aq^{-\gamma}, q \geq 4$) as a function of undamaged fraction $p$ for various values of $\gamma$ less than or equal to $1.5$. Black curves show analytic results, with the finite threshold for $\gamma=1.5$ marked by a black circle. Symbols show measurements averaged over 100 synthetic networks of $N=10^4$ nodes (circles) and $N=10^7$ nodes (squares), see Appendix \ref{simulations}.
\label{M3_small}}
\end{figure}
%%%%%%%%%%%%%%%%%%%%%%%%%%%%%%%%%%%%%%%%%%%%

To explore the region below $\gamma=2$, we must again proceed with caution. As shown in Figs. \ref{M3_small2} and \ref{M3_small}, we verify all results against numerical simulations, and find that the numerical  measurements converge to the analytical results as the system size increases.
We again introduce random damage, and use the undamaged fraction of edges or vertices $p$ as a control parameter.
The hybrid transition disappears at $\gamma = 1+1/(M-1)$. Close to (above) this point, the hybrid transition continues to exist, but the size of the jump becomes extremely small, as we show in Figure \ref{M3_small2}. A similar phenomenon was observed in mutually connected cluster \cite{baxter2012avalanche} approaching $\gamma=2$.

To find the size and location of the jump, we modify Eq. (\ref{570}) for symmetrical layers, obtaining
\begin{equation}\label{S_symmetric}
S=[1 - G(1-Z)] ^M
\end{equation}
where $Z$ obeys Eq. (\ref{580}),
so that 
\begin{equation}
S=Z ^{M/(M-1)}\label{Sbond}\,.
\end{equation}

We look for an expansion of $G(1-Z)$ for small $Z$, and keep the two leading orders,
see Eqs. (\ref{Gexp1})-(\ref{a80}) in Appendix \ref{sa2}. When $\gamma < 2$ the leading order is $\gamma-1 < 1$, see Eq. (\ref{a80}).
The self-consistency equation for $Z$ is then (for $Z \ll 1$):
\begin{equation}\label{psi}
Z \cong \left\{ -A\Gamma(1-\gamma) Z^{\gamma-1} -B Z \right\}^{M-1} \equiv \Psi(Z)\
\end{equation}
where the coefficient $B$, given by Eq. (\ref{B}), depends on the specific form of the degree distribution.

A hybrid transition occurs when the line $Z$ is tangent to $\psi(Z)$, which occurs when
\begin{equation}\label{hybrid_crit}
\left(\frac{\Psi}{Z}\right)' = \frac{1}{Z}\left[ \Psi' - \frac{\Psi}{Z}\right] = 0\,.
\end{equation}
Assuming that $Z \neq 0$ this gives us the value of $Z$ above the discontinuity:
\begin{equation}\label{Zc_general}
Z_c = \left\{
\frac{-A\Gamma(1-\gamma)  \left[(M-1)(\gamma-1)-1\right]}{B(M-2)} \right\}^{1/(2-\gamma)}\,.
\end{equation}
We see that $Z_c$ (and hence $S_c$) tends to zero at $\gamma = 1 + 1/(M-1)$.
Writing $\gamma = 1 + 1/(M-1) + \delta$, we can see that $Z_c$ near $\delta =0$ behaves as
\begin{equation}\label{Zc_Mlayers}
Z_c \cong \left[
\frac{-A(M-1)}{B(M-2)} \Gamma\left( \frac{-1}{M-1}\right)
\delta
\right]^{(M-1)/(M-2)}\,.
\end{equation}

To account for the damage applied to the network, as vertex removal, 
we use the original degree distribution, but modify Eqs. (\ref{S_symmetric}) and (\ref{580}) as follows
\begin{equation}
S=p[1 - G(1-Z)] ^M
\end{equation}
and
\begin{equation}
Z = p\left[ 1 - G(1-Z)\right]^{M-1}\,,
\end{equation}
so that 
\begin{equation}
S=p\left[\frac{Z}{p}\right] ^{M/(M-1)}\label{Ssite}\,.
\end{equation}
The self-consistency equation for $Z$ is now (for $Z \ll 1$):
\begin{equation}\label{psib}
Z \cong p\left\{ -A\Gamma(1-\gamma) Z^{\gamma-1} -B Z \right\}^{M-1} \equiv \Psi(Z)\,.
\end{equation}
Following the same procedure, applying the condition for the hybrid transition point Eq. (\ref{hybrid_crit}) gives us 
again Eq. (\ref{Zc_Mlayers}).

Substituting Eq. (\ref{Zc_Mlayers}) back into Eq. (\ref{psib}) gives %the same value for $p_c$ as before, Eq. (\ref{pc}).
\begin{multline}
p_c  \cong \!\left[-A\Gamma\!\left(\frac{-1}{\!M\!-\!1}\!\right)\right]^{-(M-1)}
\\ \times
\!\!\left[
    \frac{(M-1)}{B(M-2)} 
\delta
\right]^{-\delta(M\!-\!1)^2/(M\!-\!2)} .
\label{pc_general}
\end{multline}
Taking the limit $\delta \to 0$ the critical point tends to the constant value
\begin{align}
p_c =\!\left[-A\Gamma\!\left(\frac{-1}{\!M\!-\!1}\!\right)\right]^{-(M-1)}
\end{align}
which depends on the degree distribution only through the amplitude $A$.
We see in Fig. \ref{M3_small2} that, although the size of the discontinuity tends to zero, the critical point remains finite.
For example, for $M=3$, for the distribution used in Figs. \ref{M3_small2} and \ref{M3_small},
\begin{equation}
p_c \to \frac{1}{[A\Gamma(-0.5)]^2} \approx 0.0905... 
 \, \, \text{ as } \gamma \to 1.5^+\,.
\end{equation}
This point is marked with a black circle in Fig. \ref{M3_small}.

Finally, 
using Eq. (\ref{Ssite}) gives, for site removal:
\begin{multline}\label{Sc_Mlayers}
S_c  \cong 
 \left[
-A\Gamma\left(\frac{-1}{M-1}\right)
\right]^{2(M-1)/(M-2)}
\\ \times
\left[
\frac{  (M-1) }{ B(M-2) }\delta
\right]^{M/(M-2)}\,.
\end{multline}

\medskip

If, instead, we wish to consider edge removal, we use the amplitude $A_p = A_1p^{\gamma-1}$ as given by Eq. (\ref{a30}). Then Eq. (\ref{Zc_Mlayers}) becomes
\begin{multline}\label{Zc_Mlayersa}
Z_c \cong \left[
\frac{-A_1(M-1)}{B(M-2)} \Gamma\left(\! \frac{-1}{M-1}\!\right)
\delta
\right]^{(M-1)/(M-2)}
\\ \times
p_c^{1/(M-2)} 
\,.
\end{multline}
Combining Eq. (\ref{Zc_Mlayersa})  with Eq. (\ref{psi}), gives the same critical point, Eq. (\ref{pc_general}).

Finally, using Eq. (\ref{Sbond}) gives, for edge removal:
\begin{align}\label{Sc_Mlayers_bond}
S_c  
&\cong
\left[
\frac{  (M-1) }{ B(M-2)}\delta
\right]^{M/(M-2)}.
\end{align}

\medskip

For $1 <  \gamma < 1+1/(M-1)$ the transition is continuous, and the critical point is always zero, with extremely slow growth of the giant component as shown in Fig. \ref{M3_small}. Finite size effects are even more significant close to $\gamma=1+1/(M-1)$, but we see that measurements of finite networks approach the analytical values for $S$ as the size of the network increases.

Keeping only the leading order in Eq. (\ref{psi}) we have
\begin{equation}
Z = [-A_p\Gamma(1-\gamma)]^{M-1}Z^{(M-1)(\gamma-1)}
,
\label{600}
\end{equation}
so 
\begin{align}
Z &= [-A_p\Gamma(1-\gamma)]^{(M-1)/[1-(M-1)(\gamma-1)]} 
\nonumber\\
&\propto p^{(\gamma-1)(M-1)/[1-(M-1)(\gamma-1)]}
,
\label{610}
\end{align}
where we recalled that $A_p=A_1 p^{\gamma-1}$. The exponent $(\gamma-1)(M-1)/[1-(M-1)(\gamma-1)]$ is positive if 
$\gamma < 1 + \frac{1}{M-1}$. This gives immediately
\begin{align}
S &= [-A_p\Gamma(1{-}\gamma)]^{M/[1-(M-1)(\gamma-1)]} 
\nonumber\\
&{=} [-A_1\Gamma(1{-}\gamma)]^{M/[1-(M-1)(\gamma-1)]}p^{(\gamma-1)M/[1-(M-1)(\gamma-1)]}
.
\label{630}
\end{align}
We see that $S$ grows as a power of $p$, so $p_c = 0$ in this region. The exponent diverges as we approach $\gamma = 1 + \frac{1}{M-1}$ from below, so $S$ grows extremely slowly in this limit. This is illustrated in Fig. \ref{M3_small}. Specifically for $M=3$,
\begin{equation}
S \sim p^{3(\gamma-1)/[1-2(\gamma-1)]}.
\end{equation}
This gives $S \sim p^1$ at $\gamma=1.2$ as can be seen in the figure.

%%%%%%%%%%%%%

In the case of vertex removal (site percolation), from Eq. (\ref{psi}) we have
\begin{equation}
Z = p[-A_1\Gamma(1-\gamma)]^{M-1}Z^{(M-1)(\gamma-1)},
\label{600b}
\end{equation}
so 
\begin{align}
Z &= [-A_1\Gamma(1-\gamma)]^{(M-1)/[1-(M-1)(\gamma-1)]} p ^{1/[1-(M-1)(\gamma-1)]}.
\label{610b}
\end{align}
Hence
\begin{align}
S =& pZ^{M/(M-1)}
\nonumber \\
=& [-A_1\Gamma(1-\gamma)]^{M/[1-(M-1)(\gamma-1)]}\nonumber \\
&\times p ^{[2M-1-(M-1)^2(\gamma-1)]/\{(M-1)[1-(M-1)(\gamma-1)]\}}.
\label{630b}
\end{align}

The interval in which the hybrid transition is absent,  $1<\gamma < 1 + 1/(M-1)$, becomes increasingly small as the number of layers increases. This region vanishes as $M\to\infty$. Thus only rather fat-tailed degree distributions can maintain this singularity.

%%%%%%%%%%%%%

For the non-symmetric multiplex networks, Eqs.~(\ref{560}) and (\ref{570}), let us consider the case of $M=3$, $1<\gamma_a,\gamma_b\gamma_c<2$. Then at small $Z$ we have 
\begin{eqnarray}
Z_a &\sim& Z_b^{\gamma_b-1} Z_c^{\gamma_c-1}
,  
\nonumber
\\[3pt]
Z_b &\sim& Z_a^{\gamma_a-1} Z_c^{\gamma_c-1}
,  
\nonumber
\\[3pt]
Z_c &\sim& Z_a^{\gamma_a-1} Z_b^{\gamma_b-1}
\label{572}
\end{eqnarray}
From this system of equations we get 
\begin{equation}
Z_a \sim Z_a^{(\gamma_a-1)(\gamma_b-1)+\gamma_a\gamma_b(\gamma_c-1)/\gamma_c}
.
\label{574}
\end{equation}
The discontinuity is absent (the transition, i.e., a singularity, in this case is at zero---hyper-resilience) if the exponent of the right-hand side of Eq.~(\ref{574}) is smaller than $1$. This leads to the following condition for the absence of the discontinuity: 
\begin{eqnarray}
&&
2(\gamma_a{-}1)(\gamma_b{-}1)(\gamma_c{-}1)
\nonumber
\\[3pt]
&&
\!\!\!\!\!\!\!\!\!\!\!\!\!\!\!\!\!\! + (\gamma_a{-}1)(\gamma_b{-}1) + 
(\gamma_b{-}1)(\gamma_c{-}1) + (\gamma_c{-}1)(\gamma_a{-}1) < 1 
.
\label{576}
\end{eqnarray}
%%

%===========================================================================
%===========================================================================
%===========================================================================
%===========================================================================

\section{Conclusions \label{conclusions}}

Multi-level networks have received significant attention in recent years. The structure and resilience of such networks has generally been studied by generalizing the concepte of connected clusters to mutually connected clusters, in which there must exist a path connecting every pair of vertices in all the layers in which they participate. A node is active if it belongs to such a mutually connected cluster. This yields an  exotic percolation phase transition that is discontinuous yet retains some features of a second-order transition. The same type of transition has been found in $k$-core percolation. However, identifying the mutually connected clusters requires a global view of the multiplex network: vertices must belong to the same connected cluster in each layer.
An alternative definition of multiplex percolation was introduced in Ref. \cite{baxter2014weak}. Under this definition, a node is active if it maintains connections to active nodes in each of the layers to which it belongs. Thus the state of a node can be determined by examining the state of its neighbors. Despite this simplicity, in this paper we have shown that this "weak multiplex percolation" exhibits a complex set of critical phenomena.

When the network consists of two layers, we encounter a continuous second-order transition, as in ordinary percolation. We have shown, however that for rapidly decaying degree distributions, the giant component grows quadratically rather than linearly above the critical threshold. 
When the degree distributions of the layers are heavy-tailed, such as powerlaw distributed, we find that the giant component grows nonlinearly above the critical point, with an exponent that depends on the powerlaw decay exponent of the degree distribution. When this exponent is different in the two layers, the critical behavior depends on the smaller of the two, except when both are smaller than $2$, in which case the growth of the giant component above the critical point is determined by both powerlaw exponents.

In networks with three or more layers, the giant weak multiplex-percolation component emerges with the same discontinuous hybrid transition found in the mutually connected component. This transition consists of a discontinuity, with a square root singularity above the critical threshold. 
The weak percolation problem on an $M$-layer multiplex \ER network is equivalent
to the problem of the giant mutually connected component in the corresponding $(M-1)$-layer multiplex \ER network.
As in other network processes with this type of transition, heavy-tailed degree distributions can have a strong effect on the transition. The critical point may be reduced to zero, while the height of the discontinuity may become very small, eventually vanishing. 
Here we have shown that for weak multiplex percolation, however, at $\gamma=2$ the threshold and discontinuity are still finite. 

Exponents smaller than $2$ are not usually investigated, as the diverging mean degree means the usual locally tree-like assumptions for configuration model networks do not strictly hold. However, by carefully comparing with large scale simulations, we show that our equations give meaningful and accurate results in this regime. We show that the discontinuity and critical point
don't become zero until $\gamma = 1 + 1/(M-1)$, in an $M$ layer network. 
This differs sharply from the mutual connected component, for which $p_c=0$ at $\gamma=2$ in two layers \cite{baxter2012avalanche}, and normal percolation, where $p_c=0$ at  $\gamma=3$.
In the range of powerlaw exponents $\gamma>2$, for a large number of layers, the weak percolation behavior becomes essentially the same as that of the mutually connected component.

The weak multiplex percolation process has the advantage of being locally decidable and thus corresponds to a different type of process than mutually connected components, being defined by physical bonding to neighbors in all layers vs connectivity to all members of a cluster within each layer.
While sharing many of the same critical phenomena, the two processes differ significantly in networks with heavy-tailed degree distributions. Choosing the appropriate model for a given system is therefore important for making correct predictions about its resilience and critical behavior.

%===========================================================================
%===========================================================================
%===========================================================================
%===========================================================================

\begin{acknowledgments}
This work was developed within the scope of the project i3N, UIDB/50025/2020 and UIDP/50025/2020, financed by national funds through the FCT/MEC. This work was also supported by National Funds through FCT, I. P. Project No. IF/00726/2015 and Project No. EXPL/FIS-NAN/1275/2013. R. A. d. C. acknowledges the FCT Grants No. SFRH/BPD/123077/2016 and No. CEECIND/04697/2017.
\end{acknowledgments}
%===========================================================================
%===========================================================================
%===========================================================================
%===========================================================================

\appendix 

\section{Modification of degree distribution by removal of edges or vertices}
\label{sa1}

Let $p$ be the fraction of undeleted edges (or vertices). 
For the original, undamaged, network, $p=1$, we have the tail of the degree distribution $A_1k^{-\gamma}$ and its first moment $\langle k \rangle_1$, where the amplitude $A_{p{=}1} \equiv A_1$. 
The first moment for $p < 1$ becomes
\begin{equation}
\langle k \rangle_p = p\langle k \rangle_1 .
\label{a10}
\end{equation}

The damage acts on the scale-free degree distribution in the following way. The low-degree part of the distribution increases. In particular, (additional) vertices of degree $0$ and especially importantly ones of degree $1$ emerge. The high-degree asymptotics stays $\propto k^{-\gamma}$, but its amplitude decreases.

For establishing the relation between $A_p$ and $A_1$ for $p\ll1$, note the following. Under edge removal, for large $k$, the number of (surviving) vertices with degrees $q>k$ in the damaged network should be equal to the the number of vertices with degrees $q>k/p$ in the original network. After integration, this gives 
\begin{equation}
A_p k^{1-\gamma} \cong A_1 p^{\gamma-1}k^{1-\gamma}
.
\label{a20}
\end{equation}
So 
\begin{equation}
A_p  \cong A_1 p^{\gamma-1} 
%%< C_1
.
\label{a30}
\end{equation}

For the case of vertex removal, only a fraction $p$ of vertices survive, and these then keep each edge with probability $p$. This gives an extra factor of $p$:
\begin{equation}
\tilde{A}_p  \cong A_1 p^{\gamma} 
.
\label{a30site}
\end{equation}

%===========================================================================

\section{Generating functions}
\label{sa2}

The generating function of the degree distribution $P(q)$ is defined as 
\begin{equation}
G(x) = \sum_q P(q) x^q
. 
\label{a40}
\end{equation}
For a Poisson  degree distribution with mean degree $\langle q \rangle \equiv c$, 
\begin{align}
G(x) &= e^{-c(1-x)} , \\
G'(x) &= c e^{-c(1-x)}
. 
\label{a50}
\end{align}

For a scale-free degree distribution $P(q) = Aq^{-\gamma}$, with a minimum degree $q_0$, we can write: 
\begin{align}
G(1-x) = \sum_q Aq^{-\gamma}(1-x)^q\nonumber\\
\approx \int_{q_0}^{\infty}Aq^{-\gamma} e^{-qx} dq\,.
\end{align}
Let $y = qx$, then
\begin{align}
G(1-x) \approx A x^{\gamma-1}\int_{xq_0}^{\infty}y^{-\gamma} e^{-y} dy\,.
\end{align}
Integrating by parts twice gives 
\begin{align}\label{Gexp1}
G(1-x) &\cong 1 - \langle q \rangle x + A \Gamma(1-\gamma) x^{\gamma-1}+ \mathcal{O}(x^2).
\end{align}

The term in order $\gamma-1$ is either the leading, the second or the third term depending on the value of $\gamma$. Keeping only the leading two terms in $x$ (after the constant), we have that:
\begin{enumerate}[(i)]
\item if $\gamma>3$, then 
\begin{equation}
G(1-x) \cong 1 - \langle q \rangle x + \frac{1}{2} \langle q(q-1) \rangle x^2
, 
\label{a60}
\end{equation}
\item if $2<\gamma<3$, then
\begin{equation}
G(1-x) \cong 1 - \langle q \rangle x + A \Gamma(1-\gamma) x^{\gamma-1}
, 
\label{a70}
\end{equation}
\item if $1<\gamma<2$, 
\begin{equation}
G(1-x) \cong 1 + A \Gamma(1-\gamma) x^{\gamma-1} + Bx
\label{a80}
\end{equation}
\end{enumerate}
where the coefficient $B$ of the linear term is no longer equal to the mean degree, which diverges, but instead depends on the specific form of the distribution,
\begin{equation}\label{B}
B = -\left\{ \sum_q q[P(q) - Aq^{-\gamma}] - \zeta(\gamma-1) \right\}\,.
\end{equation}
Note that 
$\Gamma(z) < 0$ for $z \in (-1,0)$ while 
$\Gamma(z) > 0$ for $z \in (-2,-1)$.

%%%%%%%%%%%%%%%%%%%%%%%%%%%%%
%%%%%%%%%%%%%%%%%%%%%%%%%%%%%
%%%%%%%%%%%%%%%%%%%%%%%%%%%%%%%
%%%%%%%%%%%%%%%%%%%%%%%%%%%%%%%%%

\section{Numerical simulations}\label{simulations}

In this Appendix we describe the numerical procedures used in our simulations. 
To calculate each data point we use a configuration model method to generate multiple realizations of networks
 with the same degree distributions.
In the illustrative examples of Figs.~\ref{fig_M2} and~\ref{M3_large} we use the same distribution in all the layers without 
degree-degree correlations or correlations between layers.
In each layer, and in each realization, we set the degree of the nodes independently at random according to the following distribution:
\begin{align}
  P_{\textrm{SM}} (q) &{=} \frac{\left[\langle q \rangle (\gamma{-}2)\right]^{\gamma-1}}{\left(\gamma-1\right)^{\gamma-2}} 
  \frac{\Gamma\left(q{+}1{-}\gamma,\langle q \rangle [\gamma{-}2]/[\gamma{-}1]\right)}{\Gamma\left(q+1\right)}
  \label{smd1}
  \\
   &\cong \frac{\left[\langle q \rangle (\gamma-2)\right]^{\gamma-1}}{\left(\gamma-1\right)^{\gamma-2}}  q^{-\gamma} ,
\label{smd2}
\end{align}
where $\Gamma(s)=\int_0^{\infty} t^{s-1} e^{-t} dt$ is the gamma function, and $\Gamma(s,x)=\int_x^{\infty} t^{s-1} e^{-t} dt$ 
is the incomplete gamma function.
The total degree must be even, so, if the sum of all degrees is odd, we add 1 to the degree of a random node.

Each edge is shared by two nodes, so the degree can be seen as the number of `half-edges' belonging to a node.
Then, the configuration model inserts 
edges by joining uniformly at random pairs of `half-edges'.
This configuration model imposes no restrictions on the emergence of self-loops and multiple edges in the network, which, for $\gamma \leq 3$, is necessary in order for the degree-degree distribution to remain uncorrelated \cite{catanzaro2005generation}.

Equation~(\ref{smd1}) is the exact degree distribution generated by the static model for infinite $N$~\cite{catanzaro2005analytic}.  
The motivation for using this distribution is that, in the small $q$ region, it contains deviations to the asymptotic form of
 Eq.~(\ref{smd2}), which are more realistic than a pure power-law distribution.
 
We cannot, however, use the distribution of Eq.~(\ref{smd1}) to generate networks with exponent $1<\gamma\leq 2$, because 
the mean degree $\langle q \rangle$ diverges. 
Notice that our main results are obtained in terms of the amplitude $A$ and the exponent $\gamma$ of the asymptotics of 
$P(q) \cong Aq^{-\gamma}$, and, although Eq.~(\ref{smd1}) cannot describe them, there are still distributions with 
$1<\gamma\leq 2$ and finite $A$.

Figures~\ref{M3_small2} and~\ref{M3_small} show results of simulations for values of $\gamma \in (1,2]$. 
To investigate this range of $\gamma$, we generate the node's degrees from a pure power-law distribution with a minimum 
degree $q_0$:
\begin{equation}
P_{\textrm{PL}} (q) = \left\{
\begin{array}{ll}
      0 & q < q_0 ,\\
      \left(\zeta(\gamma)-\sum_{q=1}^{q_0-1} q^{-\gamma}\right)^{-1} q^{-\gamma} & q \geq q_0 ,\\
\end{array} 
\right. 
\label{pld}
\end{equation}
where $\zeta(s) = \sum_{n=1}^{\infty} n^{-\gamma}$ is the Riemann zeta function.
Notice that, unlike for Figs.~\ref{fig_M2} and~\ref{M3_large}, for Figs.~\ref{M3_small2} and~\ref{M3_small} we cannot use $\langle q \rangle$ as 
control parameter because of its divergence; 
instead we first generate networks using the distribution of Eq.~(\ref{pld}), which depends only on $\gamma$, and later 
apply damage by removing a fraction $p$ of edges at random.

For our simulations, we choose the minimum degree $q_0=4$ in Eq.~(\ref{pld}) to ensure that the transition is well 
observed at a value of $p$ smaller than $1$ for all $\gamma>1$, which means that the networks can resist some amount 
of damage before the collapse of the giant weakly percolating component.
Similar results can be obtained for any $q_0$ sufficiently large. 

In the range $1 < \gamma \leq 2$ the 
divergence of $\langle q \rangle$ leads to a dramatic increase of the amount of CPU time and memory required by simulations.
Additionally, in this extreme range of $\gamma$, there is another issue in simulations of (necessarily) finite systems, namely, a single node accumulates a large fraction of all the edges.

To elucidate this point, let us consider the effects of truncating the degree distribution at some cutoff degree $ \sim N^\alpha$ when $\gamma < 2$.
In this case, 
 the average degree is $\sim N^{\alpha(2-\gamma)}$, and the total degree of the network is $ E = N\langle q \rangle \sim N^{1+\alpha(2-\gamma)}$.
The expected number of self-loops of a node of degree $q$ is proportional to $q^2/E$, i.e., $q$ times the probability that a `half-edge’ belonging to that node is picked uniformly at random out of $E$ possibilities.
In particular, for the highest-degree node present in the system, with degree $ q_{\textrm{max}} \sim N^\alpha$, the number of self-loops is $\sim q_{\textrm{max}}^2/E \sim N^{\alpha \gamma-1}$.
For the amount of self-loops to be a vanishingly small fraction of all the edges, the ratio between the number of self-loops of the highest-degree node and the total degree $E$, which is $\sim N^{2\alpha(\gamma-1)-2}$, must go to zero as $N \to \infty$.
Then, by using an exponent of the truncation cutoff $\alpha<1/(\gamma-1)$, we can avoid the undesirable finite-size effect of a single node accumulating a finite fraction of all the edges in the form of self-loops. The same estimate can be obtained for multiple edges.
Notice that this effect only occurs for $\gamma \leq 2$, while for $\gamma > 2$ the fraction of edges that are self-loops vanishes even when we use the complete (not-truncated) degree distribution.

Since the largest value of $\alpha$ that can be used in the whole range of $\gamma < 2$ is $\alpha = 1$, in the simulations of Figs. 5 and 6 we generated the degrees from distributions truncated at $N$, i.e., $P(q<N) = P_{\textrm{PL}}(q)/\sum_{q’<N}P_{\textrm{PL}}(q’)$ and $P(q\geq N ) = 0$.
Conveniently, the use of the cutoff in the range $\gamma<2$, which avoids the explosion of self-loops, also requires much lower amounts of CPU time and memory, allowing us to explore larger system sizes $N$.

%%%%%%%%%% BIBLIOGRAPHY %%%%%%%%%%%

\bibliography{general_interdependent_networks}

\end{document}